\documentclass[aps,prl,amsmath,amssymb,floatfix,twocolumn, showpacs]{revtex4}
\usepackage{graphicx}
\usepackage{subfigure}
\usepackage{color}
\usepackage{hyperref}

\begin{document}
\title{\bf Continuous and discontinuous topological quantum phase transitions}
\author{Bitan Roy}
\affiliation{Condensed Matter Theory Center and Joint Quantum Institute, Department of Physics, University of Maryland, College Park, Maryland 20742- 4111 USA}
\author{Pallab Goswami}
\affiliation{Condensed Matter Theory Center and Joint Quantum Institute, Department of Physics, University of Maryland, College Park, Maryland 20742- 4111 USA}
\author{Jay D. Sau}
\affiliation{Condensed Matter Theory Center and Joint Quantum Institute, Department of Physics, University of Maryland, College Park, Maryland 20742- 4111 USA}
\date{\today}
\begin{abstract}
The continuous quantum phase transition between noninteracting, time-reversal symmetric topological and trivial insulators in three dimensions is described by the massless Dirac fermion. We address the stability of this quantum critical point against short range electronic interactions by using renormalization group analysis and mean field theory. For sufficiently weak interactions, we show that the nature of the direct transition remains unchanged. Beyond a critical strength of interactions we find that either (i) there is a direct first order transition between two time reversal symmetric insulators or (ii) the direct transition is eliminated by an intervening time reversal and inversion odd ``axionic" insulator. We also demonstrate the existence of an interaction driven first order quantum phase transition between topological and trivial gapped states in lower dimensions. 
\end{abstract}

\maketitle

\emph{Introduction}: The spin-orbit coupled, time reversal symmetric insulators in two and three dimensions belong to the Altland-Zirnbauer class AII~\cite{altland, ten-fold}. Due to the existence of nontrivial $Z_2$ topological invariants in both spatial dimensions, a band insulator can be classified either as a strong topological insulator (TI) or a trivial/normal insulator (NI). The four component massive Dirac fermion provides an efficient low energy description of Kramers degenerate conduction and valence bands in such systems~\cite{fu-kane,liu-zhang, TI-review-1, TI-review-2}. The sign of the Dirac mass provides the topological distinction between two insulating states. In a clean, noninteracting system the universality class of a continuous topological quantum phase transition (QPT) between them (where the Dirac mass vanishes) is described by a massless Dirac Hamiltonian. Recently, there has been considerable interest in understanding the nature of such topological transition in the two-dimensional HgCdTe quantum well~\cite{Molenkamp} and three dimensional materials such as BiTl(S$_{1-\delta}$Se$_\delta$)$_2$~\cite{Xu1, Sato} and (Bi$_{1-x}$In$_x$)$_2$Se$_3$~\cite{Brahlek, Liu}. Although the phase diagram of noninteracting, disordered systems has been investigated in many analytical and numerical works~\cite{shindou-murakami, hasting-loring, guo-franz, ostrovsky-mirlin, goswami-chakravarty, Ohtsuki-Imura, Herbut-Imura, goswami-new, roy-new} and the interplay of disorder and long range Coulomb interactions has also been studied perturbatively for both two~\cite{ostrovsky-mirlin} and three dimensions~\cite{goswami-chakravarty}, the effects of strong short range electronic interactions on the phase diagram of clean AII insulators is not well understood. The investigation of this fundamental problem is the central theme of this Rapid Communication.

\begin{figure}[htbp]
\centering
\subfigure[]{
\includegraphics[width=4.05cm, height=3.25cm]{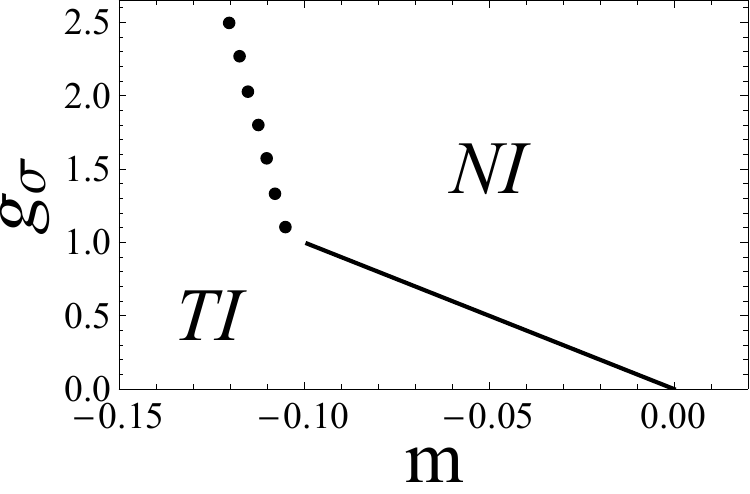}
\label{fig1a}
}
\subfigure[]{
\includegraphics[width=4.05cm, height=3.25cm]{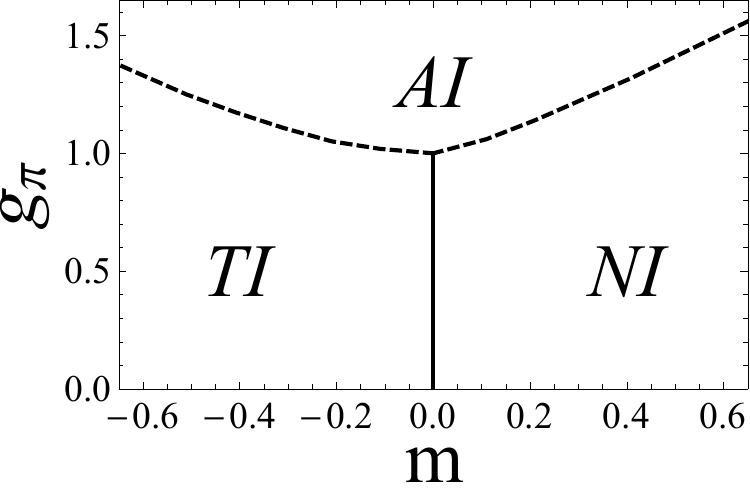}
\label{fig1b}
}
\label{phase-diagram-TI}\caption[]{Zero temperature phase diagram of three-dimensional Dirac fermion for (a) $g_\pi=0$ and (b) $g_\sigma=0$, for $b=0.2$. All the parameters are dimensionless (defined in the text). $g_\sigma$ and $g_\pi$ are interaction couplings for scalar and pseudoscalar mass generation, respectively, and $m$ is the band mass for Dirac fermions. Dirac semimetal is realized on the solid line. The transition between two insulators is first order (continuous) across the dotted (solid and dashed) line(s). The three phases TI, NI and AI meet at a multicritical point as shown in (b). For d=3, similar first order transition can occur between strong and weak topological insulators, and also between weak topological and normal insulators. }
\end{figure}

We show that the three-dimensional massless Dirac fermion is stable against sufficiently weak, but generic short range electronic interactions. Consequently, the universality class of the continuous QPT between the TI and NI remains unaffected up to a critical strength of interactions, beyond which the quantum critical point (QCP) becomes destabilized due to the spontaneous gap generation for the massless Dirac fermion. There are two possibilities for the gap: (i) an inversion  ($\mathcal{P}$) and time reversal ($\mathcal{T}$) preserving scalar mass, which does not break any \emph{bonafide} microscopic symmetry and (ii) a pseudoscalar mass that spontaneously breaks genuine discrete $\mathcal{P}$ and $\mathcal{T}$ symmetries and corresponds to an ``axionic insulator" (AI), which has recently been discussed in the context of magnetic insulators~\cite{essin, li-wang}, a Kondo singlet phase~\cite{roy-dzero}, and $ p+is$ superconductor~\cite{PG-BR}. The scalar mass generation leads to a \emph{direct first order QPT} (without band gap closer) between the TI and NI [see Fig.~\ref{fig1a}]. \emph{ This is a fluctuation driven first order transition and is distinct from the generic first order transition between two different broken symmetry phases ~\cite{raghu, pesin}}. On the other hand, the nucleation of pseudoscalar mass separates two $\mathcal{T}$ symmetric insulators, while eliminating a direct transition between them [see Fig.~\ref{fig1b}], and three phases (TI, NI and AI) meet at a multi-critical point (MCP). In contrast to the TI and NI, which are respectively characterized by the quantized magnetoelectric coefficients $\pi$ and $0$, the AI phase supports a nonquantized magnetoelectric coefficient. Consequently, the AI phase does not possess gapless surface states, and the massive fluctuations of the magnetoelectric coefficient cause dynamic magnetoelectric effects. We also reach similar conclusions regarding the fate of the QPT between the TI (quantum spin Hall insulator) and NI in two dimensions. However, the nature of the possible $\mathcal{T}$ breaking phases in two and three dimensions are considerably different. We also show a similar first order transition between topological and trivial superconductors in one dimension.

\emph{Model}: We will mostly be interested in the three-dimensional materials with $\mathcal{P}$ and $\mathcal{T}$ symmetries. For concreteness, we choose the following tight binding Hamiltonian on a cubic lattice~\cite{liu-zhang}
\begin{eqnarray}
H &=& \sum_{\mathbf{k}} \; \Psi^\dagger_{\mathbf{k}} \; [ t_1 \sum_{j=1}^{3} \Gamma_j \sin (k_j a) + \Gamma_4 \; M ]\Psi_{\mathbf{k}} \nonumber \\ &&+ t_2 \sum_{\mathbf{k}} \; \sum_{j=1}^{3} \; \Psi^\dagger_{\mathbf{k}}  \; \Gamma_4 \; \left[1 -\cos (k_j a) \right]\Psi_{\mathbf{k}},\label{TINI1}
\end{eqnarray}
as a minimal model for describing the insulating states in class AII, where $a$ is the lattice constant. The four component spinor $\Psi^\top_{\mathbf{k}}=(c_{+,\uparrow,\mathbf{k}},c_{+,\downarrow,\mathbf{k}},c_{-,\uparrow,\mathbf{k}},c_{-,\downarrow,\mathbf{k}})$ is comprised of the electron annihilation operator $c_{r,s,\mathbf{k}}$ for states with wavevector $\mathbf{k}$, the orbital parity $r=\pm$ and the spin projections $s=\uparrow/\downarrow$. The relevant discrete symmetry operations are defined according to $\mathcal{P}$: $\mathbf{k} \to - \mathbf{k}$ and $\Psi_{\mathbf{k}} \to \Gamma_4 \Psi_{-\mathbf{k}}$, and $\mathcal{T}$: $\mathbf{k} \to - \mathbf{k}$ and $\Psi_{\mathbf{k}} \to \Gamma_1 \Gamma_3 \Psi_{-\mathbf{k}}$. We have used only four mutually anticommuting matrices $\Gamma_j=\sigma_j \otimes \tau_1$ (with $j=1,2,3$), $\Gamma_4=\sigma_0 \otimes \tau_3$ in Eq.~(\ref{TINI1}) to enforce simultaneous ${\mathcal P}$ and ${\mathcal T}$ symmetries. The bilinear $\Psi^\dagger \Gamma_5 \Psi$, where $\Gamma_5=\sigma_0 \otimes \tau_2$, maintains two fold degeneracy of the valence and conduction bands, but separately breaks ${\mathcal P}$ and ${\mathcal T}$. By contrast, fermion bilinears $\Psi^\dagger \Sigma_{ab} \Psi$, with $\Sigma_{ab}=[\Gamma_a,\Gamma_b]/(2i)$, either break ${\mathcal P}$ or ${\mathcal T}$, and lift the Kramers degeneracy. Consequently, $\Psi^\dagger \Sigma_{ab} \Psi$ and $\Psi^\dagger \Gamma_5 \Psi$ are absent in $H$. Here, $\sigma_\mu$ and $\tau_\mu$ are two sets of Pauli matrices respectively operating on the spin and the parity indices. The above model describes both strong (for $-2<\frac{M}{t_2} <0$) and weak (for $-4< \frac{M}{t_2} <-2$) TIs. The strong TI and NI phases respectively correspond to $\mathrm{sgn}(M t_2)<0$ and $\mathrm{sgn}(M t_2)>0$ and $M=0$ describes the QPT between them. In contrast, the transition between weak TI and strong TI (NI) takes place when $\frac{M}{t_2}=-2 (-4)$. For simplicity, we focus on the strong TI-NI QCP. But, our results are equally applicable near the weak TI-strong TI or NI QCPs.

Since the minimal gap in the spectrum near the strong TI-NI QPT occurs at the $\Gamma$ point $\mathbf{k}=(0,0,0)$~\cite{gapcloser-comment}, the low energy quasiparticles in its vicinity are described by the following continuum Hamiltonian
\begin{equation}
H_{3d}= \int^\prime \frac{d^3k}{(2\pi)^3} \; \Psi^\dagger_{\mathbf{k}}\left[ v \; \sum_{j=1}^{3} \Gamma_j k_j + \Gamma_4(M+ Bk^2) \right]\Psi_{\mathbf{k}},\label{TINI2}
\end{equation}
where $v=t_1 a$ and $B=t_2 a^2$. The integral over the momentum is restricted to an ultraviolet cutoff $\Lambda$. The TI and NI phases respectively correspond to $\mathrm{sgn}(M B)<0$ and $\mathrm{sgn}(M B)>0$. At the QCP ($M=0$) between these two insulators the critical excitations are described by massless Dirac fermions and $B k^2$ act as a momentum dependent Wilson mass. Only the fixed point Hamiltonian of massless Dirac fermion ($M=B=0$) possesses a global chiral U(1) symmetry under which $\Psi \to \Psi e^{i \theta \Sigma_{45}}$~\cite{CS-explanation}. At the microscopic level $B \neq 0$ and $H_{3d}$ enjoys only a reduced $Z_2$ particle-hole symmetry, captured by the anticommutation relation $ \{H_{3d}, \Gamma_5\}=0$. Additionally, after an appropriate modification of the spinor, the same Hamiltonian operator can also describe the topological QPT (BEC-BCS transition) for superconducting systems in Altland-Zirnbauer classes AIII and DIII.

For simplicity, we here restrict ourselves to class AII and consider the effects of repulsive, short-range interactions. For a $\mathcal{P}$, $\mathcal{T}$ symmetric system with rotational symmetry, generic model of short-range interaction 
\begin{eqnarray}
H_{int} &=& \int d^3x \bigg [\frac{\lambda_1}{2} \left( \Psi^\dagger \Gamma_0 \Psi \right)^2 + \frac{\lambda_2}{2} \left( \Psi^\dagger \Sigma_{45} \Psi \right)^2 \nonumber \\ && + \frac{\lambda_\sigma}{2} \left( \Psi^\dagger \Gamma_4 \Psi \right)^2 + \frac{\lambda_\pi}{2} \left( \Psi^\dagger \Gamma_5 \Psi \right)^2 \bigg]
\end{eqnarray}
is described by only four linearly independent couplings due to the \emph{Fierz identity}~\cite{herbut-juricic-roy}. For $\lambda_1=\lambda_2=0$ and $M=B=0$, $H_{3d}+H_{int}$ corresponds to the celebrated Nambu-Jona-Lasinio (NJL) model for mass generation through spontaneous U(1) chiral symmetry breaking~\cite{NJL}.

\emph{RG analysis}: Since the dynamic scaling exponent for massless Dirac fermion is $z=1$, the scaling dimensions of the quartic interactions are $[\lambda_j]=(z-d)=(1-d)$. Therefore, any sufficiently weak short range interaction is an irrelevant perturbation at the massless Dirac QCP for $d>1$, and leaves the universality class of the direct TI-NI transition unchanged. In the weak coupling limit, $\lambda_i$s only modify the phase boundary in a non-universal manner. These features and the potential breakdown of the massless Dirac theory for strong interactions can be captured through a renormalization group (RG) analysis, controlled via simultaneous $1/N_f$- and $\epsilon$-expansion about the \emph{lower critical dimension} $d=1$, where $N_f$ is the flavor number of four component fermions and $\epsilon=d-1$. Within the Wilsonian momentum shell method, we integrate out the degrees of freedom inside $-\infty < \omega < \infty$ and $\Lambda e^{-l} < k< \Lambda$, and subsequently rescale according to $x \to x e^l$, $\tau \to \tau e^l$, $\Psi \to e^{-d l/2} \Psi$. In the $N_f \to \infty$ limit, we obtain the following flow equations
\begin{eqnarray}\label{RGintcoup}
\frac{dm}{dl} = m \left(1+ 2 g_\sigma \right)+ 2 g_\sigma b, \: \: \frac{db}{dl}=-b, \: \:
\frac{dg_1}{dl} = -\epsilon g_1, \: \: \:
 \nonumber \\
\frac{dg_2}{dl} = -\epsilon g_2,
\frac{dg_\sigma}{dl} = -\epsilon g_\sigma + 2 g^2_\sigma , \: \:
\frac{dg_\pi}{dl} = -\epsilon g_\pi + 2 g^2_\pi,
\end{eqnarray}
where we have introduced the dimensionless couplings $g_i=\lambda_i S_d N_f \Lambda^{d-1}/[v(2\pi)^d]$, $m=M/(v \Lambda)$, $b=B\Lambda/v$, and $S_d$ is the surface area of a $d$-dimensional unit sphere.

For sufficiently weak interactions, Dirac mass $m$ is the only relevant variable. In this regime, $g_i (l) \sim g_{i,0} e^{-\epsilon l}$, $b(l) \sim b_0 e^{-l}$, where $g_{i,0}$ and $b_0$ are the bare values of the corresponding couplings and the solution of Eq.~(\ref{RGintcoup}) yields
\begin{equation}\label{mass-RG}
m(l)+ \frac{2}{d+1} b(l) g_\sigma(l)=e^l \left(m_0+\frac{2}{d+1} b_0 g_{\sigma,0} \right).
\end{equation}
Therefore, $m^\ast= m_0+\frac{2 \; b_0}{d+1} g_{\sigma,0}$ acts as an effective mass and the TI-NI phase boundary is determined by $m^\ast=0$. This agrees well with the phase boundary determined through minimizing the free energy for weak interactions, as shown in Fig.~\ref{fig1a} for $d=3$~\cite{flavornumber}. The flow equations also show that $g_1$ and $g_2$ are always irrelevant perturbations, and can be ignored. Thus, we can restrict ourselves to the NJL model that supports three unstable fixed points: (i) $g^c_\sigma=\epsilon/2$, $g^c_\pi=0$; (ii) $g^c_\sigma=0$, $g^c_\pi=\epsilon/2$; and (iii) $g^c_\sigma=g^c_\pi=\epsilon/2$. The first two are QCPs, respectively describing the generation of scalar and pseudoscalar masses for massless Dirac fermion at strong couplings, whereas the third one is a \emph{bicritical} point. For $d>1$ and at strong coupling ($g_i>g^{c}_i$), the appropriate correlation length $\xi_i \sim \Lambda^{-1} (g_i -g^c_i)^{-\nu_i}$ provides the infra-red cutoff for the RG flow, with a correlation length exponent $\nu_i=1/\epsilon$. Therefore, for $d=2$ and $d=3$ we respectively have non-Gaussian and Gaussian (mean-field) itinerant QCPs~\cite{moshe-moshe}. As $b(l)=b_0 e^{-l}$, we can ignore it in the RG sense in the weak-coupling regime. However, the effects of $b$ cannot be neglected for $g_i>g^{c}_i$s. For elucidating the dramatic effects of $b$ in determining the nature of the strong coupling phases and the associated transitions, next we adopt the method of large $N_f$ mean-field theory.

\emph{Free energy}: We perform the Hubbard-Stratonovich decoupling of the quartic terms proportional to $g_\sigma$ and $g_\pi$, with bosonic fields $\Sigma$ and $\Pi$, respectively, that couple to the fermion bilinears as $\Sigma \; \Psi^\dagger \Gamma_4 \Psi$ and $\Pi \; \Psi^\dagger \Gamma_5 \Psi$. When the corresponding coupling constants exceed their critical strengths, these bosonic fields acquire expectation values, while giving rise to the gap in the spectrum. With $\langle \Pi \rangle =0$, $\langle \Sigma \rangle \neq 0$, $\langle \Sigma \rangle + M$ acts as the effective scalar mass, with TI and NI being the only possible phases. In contrast, when $\langle \Pi \rangle \neq 0$, $\mathcal{P}$ and $\mathcal{T}$ are spontaneously broken, giving rise to the AI phase. Since the free energy density ($F$) has the dimension $E L^{-d} \sim L^{-(d+1)}$, we define a dimensionless quantity $f=F (2\pi)^d /(N_f S_d v \Lambda^{d+1})$, which takes the form
\begin{equation}
f=\frac{\sigma^2}{2 g_\sigma}+\frac{\pi^2}{2 g_\pi} -2 \int^{1}_0 dx x^{d-1} \sqrt{x^2+(\sigma+m+b x^2)^2+\pi^2}.
\end{equation}
We have also introduced the following dimensionless variables $\sigma=\Sigma/(v \Lambda)$, $\pi=\Pi/(v\Lambda)$, $x=k/\Lambda$.

Note $\sigma=-m$, $\pi=0$ corresponds to the massless Dirac fermion describing the continuous QPT between the TI and NI, and remains as the global minimum of $f$ up to critical strengths of $g_i$s. Consequently, the phase boundary between the TI and NI is determined by
\begin{equation}
m=\frac{2 g_\sigma}{3 b^3}\left[(2-b^2) \sqrt{1+b^2} -2\right],
\end{equation}
in $d=3$, which is equivalent to $m^\ast=0$ for small $b$, up to the critical strengths of the quartic interactions
\begin{eqnarray}
g^\ast_\sigma =\frac{1}{2}\left[1+b^2+ \sqrt{1+b^2} \right], \:
g^\ast_\pi = \frac{1}{2} \left[1+ \sqrt{1+b^2} \right].
\end{eqnarray}
Notice that $g^\ast_\sigma > g^\ast_\pi$ for an arbitrary value of $b$. This is related to the fact $\{ \Gamma_4, \Gamma_5 \}=0$. Thus a finite $b$ favors the nucleation of the pseudoscalar mass. In a continuum model without any momentum dependent mass ($b=0$) the critical couplings $g^\ast_\sigma=g^\ast_\pi=\epsilon/2=1$ become identical to the ones found from the RG analysis of the NJL model.

We numerically minimize the free energy to obtain the phase diagrams in Figs.~\ref{fig1a} and ~\ref{fig1b}. However, all the salient features can be understood by expanding $f$  for small $\sigma+m$ and $\pi$, and retaining only the lowest order $b$-linear contributions. The free energy is then given by
\begin{align}
f_{3d} \approx \left[ \frac{m^2}{2g_\sigma}-\frac{1}{2}\right]-\sigma \left[ \frac{m}{g_\sigma}+\frac{b}{2} \right]+ \frac{\sigma^2}{2} \left[ \frac{1}{g_\sigma}-\frac{1}{g^c_\sigma} \right] + \frac{\pi^2}{2} \nonumber \\
\left[ \frac{1}{g_\pi}-\frac{1}{g^c_\pi} \right]
+ \frac{b}{2} \sigma (\sigma^2+\pi^2) + \frac{(\sigma^2+\pi^2)^2}{4} \log \left[ \frac{2 e^{-1/4}}{\sigma^2+\pi^2} \right], \label{freeen3d}
\end{align}
for $d=3$, after shifting $\sigma \to \sigma -m$. When we consider the nucleation of the scalar mass (for sufficiently strong $g_\sigma$), the following subtleties need to be addressed: (i) $\left( \frac{m}{g_\sigma}+\frac{b}{2} \right)$ acts as the external field for $\sigma$ and (ii) the momentum dependent mass $\propto b$ gives rise to all odd powers (cubic and higher) of $\sigma$ in the expression for the free energy density. \emph{Due to the presence of all the odd powers of $\sigma$, the nucleation of the scalar mass proceeds through a first order QPT between the TI and NI [see Fig.~\ref{fig1a}]}. By contrast, the pseudoscalar mass generation (for sufficiently strong $g_\pi$) occurs through continuous QPT. Three phases (TI, NI and AI) meet at a MCP (located at $m^\ast=0$ and $g_\pi=g^\ast_\pi$ in the $g_\sigma =0$ plane), as shown in Fig.~\ref{fig1b}. The MCP displays mean-field exponents, but the hyperscaling is violated by logarithmic corrections [due to the $\pi^4 \log(\pi^2)$ term in Eq.~(\ref{freeen3d})].

\emph{Two-dimensions}: Our study can be generalized to address the transition between the TI (quantum spin-Hall insulator) and NI in two dimensions. In the continuum limit, these two insulators and the transition between them can be described by the Bernevig-Hughes-Zhang model~\cite{BHZ}
\begin{align}
H_{2d} = \int^\prime \frac{d^2k}{(2\pi)^2} \Psi^\dagger_{\mathbf{k}}\big[ v \left( \Gamma_3 k_x + \Gamma_5 k_y \right)
+ \Gamma_4 (M+ Bk^2) \big]\Psi_{\mathbf{k}}.
\label{BHJ}
\end{align} 
The TI and NI phases in this model are realized for $\mathrm{sgn}(MB)<0$ and $\mathrm{sgn}(MB)>0$, and the continuous QPT between them ($M=0$) is described by two-dimensional massless Dirac fermion. Strong  interactions can give rise to the following four Dirac masses: $\Sigma=\langle \Psi^\dagger \Gamma_4 \Psi \rangle$, $\Pi_{1,2}=\langle \Psi^\dagger \Gamma_{1,2} \Psi \rangle$, and $\Xi=\langle \Psi^\dagger \Sigma_{35} \Psi \rangle$. As in $d=3$, the nucleation of the scalar mass $\Sigma$ (for sufficiently strong $g_\sigma$) proceeds through a first order QPT, which can be understood from the following free energy density 
\begin{eqnarray}
f_{2d} &=& \left( \frac{m^2}{2g_\sigma}-\frac{2}{3}\right)-\sigma \left( \frac{m}{g_\sigma}+\frac{2}{3}b\right)+\frac{\sigma^2}{2} \left( \frac{1}{g_\sigma}-\frac{1}{g^c_\sigma}\right) \nonumber \\
&+& |\sigma|^3 \left(\frac{2}{3}+b \; \text{sgn}(\sigma) \right) + {\cal O} (b^2, \sigma^4),
\end{eqnarray}
 where $g^c_\sigma=1/2$. The corresponding phase diagram is qualitatively similar to Fig.~\ref{fig1a}. In contrast, condensation of $\mathcal{T}$-odd magnetic masses $\Pi_{1,2}$  or the anomalous charge Hall mass $\Xi $~\cite{haldane} takes place through a continuous QPT as in Fig.~\ref{fig1b} (see also Ref.~\cite{pikulin}), giving rise to a MCP in the $(m,g_\pi)$ plane (with non-mean-field exponents~\cite{HJR-O2, assaad}). Our proposed first order transition between TI and NI has recently been observed in numerical works in both $d=2$~\cite{amaricci-1} and $d=3$~\cite{amaricci-2}.

\emph{One dimension}: \emph{In one dimension, the quartic interaction ($g_\sigma$) is marginally relevant and destabilizes the massless Dirac fixed point for infinitesimal strength} [setting $\epsilon=0$ in Eq.~(\ref{RGintcoup})]~\cite{gross}. In the presence of a momentum dependent mass, the QPT between topological and trivial gapped states can be first order for a sufficiently large number of flavors. This is the reason behind the existence of a fluctuation driven first order QPT for the $N$-color Ashkin-Teller chain when $N \geq 3$~\cite{widom, fradkin, shankar}. The pertinent continuum model is described by $N$ species of two component Majorana/Jordan-Wigner fermions with a $k$ dependent mass in the vicinity of the decoupled Ising QCP, and microscopic four-spin coupling gives rise to $O(N)$ invariant quartic interaction. Such a first order QPT can be germane for describing the direct QPT between topological and trivial superconductors in different classes.

\emph{Conclusions}: Our results can also shed light onto the finite temperature phase diagram of strongly interacting Dirac fermion. At finite temperatures and for  sufficiently strong interactions, the massless Dirac fermion can undergo either (i) a first-order classical phase transition and enter into the TI or NI phase, or (ii) a continuous classical phase transition, giving rise to a $\mathcal{T}$-breaking insulator. In general, the semiconductors (where most of the TIs have been identified) are weakly correlated materials. However sufficiently strong electronic interactions (local and nonretarded) can be mediated by optical phonons, below the scale of optical frequency~\cite{maki, gruner, kivelson}.

Finally, we comment on the effects of disorder on various QPTs, discussed for clean systems. Harris criterion dictates that any continuous QPT is stable against weak disorder if $\nu >2/d$~\cite{Harris} and this has been explicitly shown for the TI-NI QPT at $d=3$~\cite{goswami-chakravarty}. Since $\nu=1/2<2/3$ at the clean MCP in $d=3$, the universality class will be changed by an infinitesimal amount of randomness into a disorder controlled class satisfying $\nu>2/3$~\cite{Spencer}. At present it is not well established whether the exact value of $\nu$ for MCP in $d=2$ is bigger or less than one ($=2/d$), and we cannot properly assess its stability against weak disorder. On the other hand, in the thermodynamic limit, the first order transition (both classical and quantum) will be rounded by disorder into a continuous one in $d=1,2$, while it can survive in $d=3$ for sufficiently weak randomness~\cite{Imry, Hui, Wehr-1, Wehr-2, Chakravarty, Greenblatt, Vojta}. The analysis of the universality class at a putative disorder and interaction controlled critical point is beyond the scope of the present Rapid Communication.

\emph{Acknowledgements:} B. R. and J. D. S. were supported by the start-up grant of J. D. S. from the University of Maryland. P. G. was supported by NSF-JQI-PFC and and LPS-CMTC.



\begin{thebibliography}{}
\bibitem{gapcloser-comment} The QPT between weak TI and strong TI (NI) is described by three copies of four component massless Dirac fermion at $X (R)$ points of the Brillouin zone.
\bibitem{CS-explanation} Since $B$ is an irrelevant parameter and can be neglected with respect to the $k$-linear term for small $k$, massless Dirac fermion at the TI-NI QCP possesses a U(1) chiral symmetry.
\bibitem{flavornumber} For TI-NI QPT, described by the model in Eq.~(\ref{TINI1}), $N_f=1$. Systematic inclusion of $1/N_f$ corrections only changes the shape of the phase boundaries. However, nature of QPTs among topologically distinct insulators is insensitive to the actual value of $N_f$.
\end{thebibliography}
\end{document}